\documentclass[pre,preprint]{revtex4}
\usepackage{psfig}

\begin{document}

\title{Role of bulk and of interface contacts in the behaviour of model dimeric proteins.}
\author{G. Tiana$^{1,2}$, D. Provasi$^{1,2}$ and R. A. Broglia$^{1,2,3}$}
\address{$^1$Department of Physics, University of Milano,}
\address{via Celoria 16, 20133 Milano, Italy,}
\address{$^2$INFN, Sez. di Milano, Milano, Italy,}
\address{$^3$The Niels Bohr Institute, University of Copenhagen,}
\address{Bledgamsvej 17, 2100 Copenhagen, Denmark}
\date{\today}

\begin{abstract}
Some dimeric proteins first fold and then dimerize (three--state dimers) while others first dimerize and then fold (two--state dimers).
Within the framework of a minimal lattice model, we can distinguish between sequences obeying to one or to the other mechanism on the basis of the partition of the ground state energy between bulk than for interface contacts. The topology of contacts is very different for the bulk than for the interface: while the bulk displays a rich network of interactions, the dimer interface is built up a set of essentially independent contacts. Consequently, the two sets of interactions play very different roles both in the the folding and in the evolutionary history of the protein. Three--state dimers, where a large fraction of the energy is concentrated in few contacts buried in the bulk, and where the relative contact energy of interface contacts is considerably smaller than that associated with bulk contacts, fold according to a hierarchycal pathway controlled by local elementary structures, as also happens in the folding of single--domain monomeric proteins. On the other hand, two--state dimers display a relative contact energy of interface contacts which is larger than the corresponding quantity associated with the bulk. In this case, the assembly of the interface stabilizes the system and lead the two chains to fold. The specific properties of three--state dimers acquired through evolution are expected to be more robust than those of two--state dimers, a fact which has consequences on proteins connected with viral diseases.
\end{abstract}

\maketitle

\section{Introduction}

Dimers are a rather common structure adopted by proteins to perform their biological
activity. They are proteins whose native conformation 
is a globule build out of two disjoint chains. In particular, homodimers are
dimers composed of two identical sequences.

The importance of the study of the evolutionary properties of homodimers
is connected with the fact that enzymes of this type produced by viral
agents (e.g. HIV--1--protease) are able to display very fast
evolutionary patterns to escape from the pressure
excerted by the immune system and by drugs. The knowledge of the
evolutionary properties of dimers can thus be of help in designing
a strategy to deal with the associated diseases.

Some of the known homodimers, like {\it E. Coli} Trp Repressor \cite{tsai}, fold
according to a three--state mechanism, where first the denaturated chains of the monomers assume 
conformations rich of native structures independently of each
other, and subsequently the two parts come together to form the dimer. 
A different behaviour is
displayed by, for example, P22 Arc Repressor, whose chains dimerize without 
populating any monomeric native--like
intermediate (two--state process). 
In this case one can only detect the unfolded monomers and the native dimers \cite{ric}.
A class by itself is composed by dimers which have
evolved
by swapping entire domains of each monomer ("domain swapping dimers", \cite{eisenberg}) so that
a segment of a monomer is replaced by the same segment of the other monomer.

Making use of a simple model of protein folding, it has been previously shown \cite{dimero_prot},  
that the folding of dimeric proteins
is, to a large extent, controlled by their ground state energy,
as already found for monomeric proteins. In the case of monomers, good folders are those
sequences which display a low energy in the native state (i.e. an energy below a threshold $E_c$, quantity
determined mainly by the length of the chain and by the statistical properties of the 
interaction matrix \cite{sh_prl,jchemphys3}). This is also found to be true for dimeric
sequences. Moreover, in this case one can distinguish between two-- and three--state dimers by
studying how the relative contact energy\footnote{Because the bulk energy is always much larger than that associated with the interface, due to the fact that the number of native bulk contacts is much larger than the number of interface contacts, we shall refer, as a rule, to relative bulk and relative interface contact energies.} is partitioned between the bulk (i.e. the interactions between
residues belonging to the same monomer divided by the corresponding number of contacts) and the interface (i.e. the interactions between
residues belonging to different monomers divided by the number of interface contacts). 

From the evolutionary point of view, it is possible to obtain one or the other behaviour concentrating the evolutionary pressure either on the bulk or on the
interface amino acids.
If the evolutionary pressure, and consequently the stabilization energy per contact, is concentrated on the bulk, the dimer displays a three--state folding mechanism. 
The resulting behaviour is, in this case, similar to that displayed by monomeric chains, where local elementary structures (LES), formed and stabilized at early stages of the folding process, essentially control the folding process. The assembly of the LES into the folding nucleus (FN) allows the chain to overcome the main free energy barrier in the path
towards the monomeric native state \cite{jchemphys3}. For three--state dimers there is an additional step in the
folding hierarchy, corresponding to the assembly of the two folded monomers into the native dimer. On the other hand,
if the overall relative contact energy is low but the relative energetic balance favours the interface contacts, a two--state dimerization process takes place, where first the interface is built and then the other residues fold around it. In this case a larger number of residues than in the case of three--state dimers are involved in the folding process, essentially all those belonging to the interface, rather than just few, highly conserved, strongly interacting, residues as in the case of three--state folders.

 The two cases, however, are not symmetric. The bulk of the two dimers
contains a rich network of interactions, while the interface is kept together by
bonds which are, to a large extent independent on each other. This topological difference gives rise to 
different evolutionary properties for the residues at the interface with respect to those buried
inside the monomers.
The evolutionary features of model dimers can be compared with those of real dimers through
the analysis of the conservation patterns in families of analogous proteins. 

The model we use to study homodimers is largely employed in the literature 
\cite{go,dill,sh_prl}
because, in spite of the simplifications introduced in the description of the protein so as to make the calculations feasable,
it still contains the two main ingredients which are at the basis of the
distinctive properties of proteins: polymeric structure and disordered interactions
among the amino acids\cite{wolynes}.
According to this model, a protein is a chain of beads on a cubic lattice, each bead representing
an amino acid which interacts with its nearest neighbours through a contact energy as described
in ref. \cite{mj} (for details about the model see, e.g., refs. \cite{jchemphys1,sh_prl}).

\section{Selecting folding sequences by reproducing evolution.}

In the case of monomeric sequences, the canonical ensemble associated with the space of sequences
for a fixed native conformation has been proved useful \cite{sh_design} in selecting proteins 
with given thermodynamical and kinetical properties, due to the fact that these properties are 
essentially determined by the native (ground state) energy.
In this context, the energy of a sequence is controlled by an "evolutionary" temperature 
$\tau$, in such a way that the probability of selecting a sequence with energy $E$ is
proportional to $\exp(-E/\tau)$. $\tau$ 
is an intensive variable which plays the role of temperature 
and gives the degree of bias towards low--energy sequences.
In the evolutionary context, $\tau$ has the meaning of selective pressure with respect to
the protein ability to fold \cite{isles}.
In particular, for values of $\tau$ lower than
the temperature $\tau^c\equiv (\partial S/\partial E|_{E=E_c})^{-1}$ the average energy of
the selected sequences is lower than $E_c$, which is the energy which separates sequences
with a unique and stable native conformation and able to find it rapidly from random heteropolymers, which undergo a lengthy and nonunique compaction process \cite{sh_design,jchemphys3}.

In the case of homodimers, we want not only to distinguish good from bad folders, but also
to select those displaying a two--state from those displaying a three--state folding mechanism.
This is done by controlling separately the energy $E_1$ associated with the interaction
of amino acids belonging to the same monomer and the energy $E_t$ associated with the interaction
of amino acids belonging to different monomers. Accordingly, we select sequences $\{s\}$ with probability
\begin{equation}
\label{dis_dim}
p(\{s\})=\frac{1}{Z}\exp\left(-\frac{H_1}{\tau_1}-\frac{H_t}{\tau_t}\right), 
\end{equation}
where $H_1$ and $H_t$ are, respectively, the bulk and the interface 
energy of the sequence on the chosen
native conformation and $Z=\sum_{\{s\}}\exp(-H_1/\tau_1-H_t/\tau_t)$
is the normalization constant. The parameters $\tau_1$ and $\tau_t$ are
intensive variable which give the degree of evolutionary pressure on the bulk and on the 
interface. 

The non--equilibrium distribution $p(\{s\})$ of Eq. (\ref{dis_dim}) is the stationary
distribution which maximizes the Shannon entropy of the system at given values of the
average bulk and interface energies. In other words, we are interested in a distribution
which does not depend on time and which forgets about all microscopic details of the
system (i.e. minimizing the information, that is maximizing the entropy), except for the two average energies over which we
want to have control. Of course, if the two average energies are set to different values,
this is a non--equilibrium distribution, implying that each of the two parts of the system is
in contact with its own thermal bath. 

Anyway, it is possible to describe the system in a fashion which is
formally similar to that of the equilibrium canonical ensemble.
If we call $p$ a generic distribution of states of the system, 
we can define the average energy functional $E_1[p]=\sum H_1 p$ and 
$E_t[p]=\sum H_t p$ of the two parts of the system and the
entropy functional $S[p]=-\sum p\log p$, which indicates the 
information we have about the system.

Among all possible distributions $p$, we are interested in the 
stationary distribution $p^*$ which minimizes the information
(maximizing $S$) at fixed values of the average energy $E_1[p]$ and $E_t[p]$, values
which we call $E_1^*$ and $E_t^*$, respectively.
In keeping also with the constrain $\sum p=1$, one minimizes the functional
\begin{equation}
S[p]-\alpha(E_1[p]-E_1^*)-\beta(E_t[p]-E_t^*)-(\log Z+1)(\sum p-1),
\end{equation}
where $\alpha$, $\beta$ and $\lambda$ are the Lagrange multipliers which
fix the value of the average energies and of the normalization factor, respectively.
Setting the derivative of this functional to zero gives the seeked expression for the
stationary probability, that is,
\begin{equation}
p^*=\frac{1}{Z}\exp(-\alpha H_1-\beta H_t),
\end{equation}
where the evaluation of the Lagrange multipliers gives $Z=\sum\exp(-\alpha H_1-\beta H_t)$, 
which has the form of a partition function, and
\begin{eqnarray}
\left.\frac{\partial S}{\partial E_1}\right|_{E_1^*}=\alpha; && \left.\frac{\partial S}{\partial E_t}\right|_{E_t^*}=\beta.
\end{eqnarray}
In parallel with equilibrium thermodynamics, we call "temperature" the
inverse of the two Lagrange multipliers, $\tau_1=\alpha^{-1}$ and $\tau_t=\beta^{-1}$. 
The simultaneous action of the two evolutionary temperatures induce, in the design of dimers, a selective bias towards sequences displaying a conspicuous low energy in the native conformation, in a similar way that a single evolutionary temperature does in the case of the design of single--domain monomeric proteins \cite{jchemphys3}. Lowering $\tau_1$ or $\tau_t$ increases the pressure set in the bulk or on the interface, respectively.

To implement this procedure we use a multicanonical technique \cite{multic,borg} in the space
of sequences, for a fixed dimeric native conformation. 
First, we select a target conformation built of two identical parts (in the present case each a 36--mer) having a face in contact,
in such a way that the overall structure is symmetrical with respect to the interface\footnote{Whether the monomers are identical or related by mirror simmetry is immaterial within the framework of the minimal model of protein folding we employ.}, as e.g.
in Fig. 1(a), choosing a realistic ratio among the different kinds of amino acids \cite{creighton}.We swap amino acids, thus keeping the "wild--type" concentration fixed, accepting or rejecting
the swap with the help of a multicanonical algorithm. 
In this way one can
select a set (composed typically of $10^4$ elements) of evolutionary uncorrelated sequences $\{s\}$ corresponding to a given pair of values $\tau_1$ and $\tau_t$, thus all designed under similar evolutionary conditions.

\section{Properties of the space of sequences}

The dynamical simulation of the folding process of sequences selected at different values of $\tau_1$ and $\tau_t$ produces the phase diagram displayed in Fig. 2, testifying to the fact that the folding properties of a sequence depend on its ground state energy. One can identify four areas in the phase diagram, corresponding respectively to a two-- or to a three--state folding behaviour and to two different kinds of aggregations: specific (i.e. depending on certain contacts) or not (i.e. a random collapse).

Sequences selected at very low \footnote{That is, low
with respect to the only energy scale of the system, which is the standard deviation
of the interaction matrix $\sigma=0.3$.} values of $\tau_1$ and $\tau_t$ fold
according to a two--state mechanism, first dimerizing and then folding to their
dimeric native state. Sequences following this behaviour are indicated with solid
squares in Fig. 2.
Rising $\tau_t$ produces sequences which fold according to a three-state mechanism,
first folding to the monomeric native state and then dimerizing (solid triangles). 
Leaving fixed $\tau_t$ we find that the upper limit of $\tau_1$ still leading to designed sequences folding according to any of these two paradigms corresponds to that value for which the total energy $2E_1+E_t=E_c$ ($\approx-35$ in the units we are considering, $RT_{room}=0.6kcal/mol$). Examples of sequences folding according to the two mechanisms are
\begin{eqnarray}
S_{1}&\equiv&\text{\tiny VLNLGNFVGGHCRYDMEASLWTAKPKPTIRISEADQ (two--state),} \nonumber\\
S_{3}&\equiv&\text{\tiny NTKPVERNCTRVIDGDFALYSGAGSMKLQEHLWPIA (three--state),} \nonumber
\end{eqnarray}
whose energies are $E_{designed}(=2E_1+E_t)=-36.1$ $(E_1=-15.50\,,E_t=-5.11)$ and $E_{designed}=-36.8$ $(E_1=-16.47\,,E_t=-3.89)$, respectively.

Keeping $\tau_t$ low and increasing $\tau_1$ (so as to meet the curve associated with
$E_c$) leads to sequences which aggregate (empty diamonds in Fig. 2). An example of such sequences being provided by
\begin{eqnarray}
S_{5}&\equiv&\text{\tiny CGNLVNGHVFGLASMKPRPSDIQWTREAIODYELTA (aggregation),}\nonumber
\end{eqnarray}
with $E_{designed}=-35.0$ $(E_1=-14.95\,,E_t=-5.11)$

Aggregation takes place because the interface becomes too reactive with respect to the bulk. 
For evolutionary temperatures outside the area defined by the ordinate ($\tau_1=0$) and $E_c$ (solid curve
in Fig. 2), the equilibrium state is, at any temperature,  
either a disordered clump made by the two chains, or a state where the two
chains are separated and unfolded, depending on their concentration (cf. 
Section V). 

The central role played by the relative contact energy associated with the bulk and with the interface can be assessed from Fig. 3. Due to the fact that typical homodimers display many more bulk than interface contacts (e.g. in the case under discussion 80 and 12 respectively, cf. Figs. 1(b) and 1(c), respectively), it is not very useful to compare the total values $2E_1$ and $E_t$, but rather the average values $\epsilon_1=2E_1/(\#\; of\; bulk\; contacts)$, $\epsilon_t=E_t/(\#\; of\; interface\; contacts)$, $\epsilon_{designed}=E_{designed}/(total\; \#\; of\; contacts)$ and $\epsilon_c=E_c/(total\; \#\; of\; contacts)$.

Fig. 3 suggests that provided $\epsilon_{designed},\epsilon_1<\epsilon_c$ the system folds. As a three--state dimer if $\epsilon_1<\epsilon_t$, and as a two--state dimer if $\epsilon_1>\epsilon_t$. It is furthermore shown that two--state dimers are less stable than three--state dimers ($\epsilon_{designed}(3-state)<\epsilon_{designed}(2-state)$), which in turn are again less stable than the monomer ($\epsilon(S^{36})<\epsilon_{designed}(3-state)$) which folds to the native conformation corresponding to one of the two identical halves which build the dimer (cf. Fig. 1(a)).

Since the folding properties of homodimeric sequences depend on their 
conformational ground state energy, it is interesting to study the 
energy landscape of the space of sequences
for a given conformation, space which is responsible for the evolution
of the corresponding dimer. 
For this purpose, we have calculated
the entropy as a function of the
bulk energy $E_1$ and of the interface energy $E_t$ in the space of sequences, making use of a
multicanonical algorithm \cite{multic,borg}, keeping fixed the conformation 
of Fig. 1. The results are displayed in Fig. 4. Also displayed in this figure are the absolute minima for the bulk and for the interface contacts, respectively

As already seen from Fig. 3, it is not possible to optimize
the bulk and the interface at the same time. In fact, lowering $\epsilon_t$ is done at the expenses of $\epsilon_1$ and vice versa. Furthermore, the results displayed in Fig. 4 indicate that the condition
of energy minimum for the bulk energy ($E_1^{min}=-17.1$) implies that the interface
energy is quite far from its minimum ($E_t=-3.4$). Vice versa, the minimum of the
interface energy is $E_t^{min}=-5.11$ is associated with a bulk energy $E_1=-16.2$. 
In keeping with this result, it is seen that small changes of $E_1$ are correlated with large variations in $E_t$, but not vice versa. In fact, a $2.5\%$ change in $E_1$ is correlated with a $30\%$ change in $E_t$, much larger than the ratio between the number of bulk and interface contacts ($\approx 7$). In other words, bulk contacts play a central role in the design of both 3--step as well as 2--step folding dimers, as testified by the fact that in both cases $\epsilon_1<\epsilon_c$, a condition not required to be fulfilled by $\epsilon_t$.

The exclusion relation observed between a simultaneous low value of $E_1$ and of $E_t$ results in  negative specific heats $C_{1t}\equiv \partial E_1/\partial \tau_t$ and $C_{t1}\equiv \partial E_t/\partial \tau_1$
at low values of $\tau_t$ and $\tau_1$, respectively, as testified by the decrease of the energy when increasing the temperature (cf. Figs. 5(a) and 5(b)). The regions of the $(\tau_1,\tau_t)$--plane corresponding to negative specific heats 
$C_{1t}$ and $C_{t1}$ are delimited by dashed curves in Fig. 2. Two--state folding sequences
lie across the line $C_{1t}=0$, meaning that a low value of $\tau_t$ is not enough
to guarantee a two--state folding character, but also $\epsilon_1$ has to be
low (cf. Fig. 3) Note that $C_{1t}=0$ implies that $E_1$ is a minimum as a function of $\tau_t$. In fact,
sequences which are in the low--$\tau_t$ region but are in the negative--$C_{1t}$ region
aggregate (empty diamonds in Fig. 2), since the associated values of $E_1$ are high (cf. Fig. 3, sequence $\#5$).
On the other hand, three-state folders are quite insensitive to the value of $E_t$ and
consequently can be found equally well on both sides of the line corresponding to 
$C_{t1}=0$ (i.e. also $\epsilon_t>\epsilon_c$ in Fig. 3 and solid traingles in Fig. 2).

Another interesting feature of the space of dimeric sequences is that the bulk energy
$E_1$ depends strongly on $\tau_1$ (the specific heat $C_{11}\equiv \partial E_1/\partial \tau_1$
ranging from $2$ to $10$) but weakly on $\tau_t$, the derivative $C_{1t}$ being
approximately a constant around $0.2$ (except, of course, in the low--tail of $\tau_t$
where $C_{1t}$ is negative). This can also be seen in Fig. 3 from the fact that a $30\%$ change in $\epsilon_t$ (in going from sequence $\#3$ to sequence $\#1$) is related to a modest change in $\epsilon_1$. On the other hand, the interface energy $E_t$, athough depending strongly only on $\tau_t$, is more sensitive
to both evolutionary temperatures, as it is clear from the fact that  
$C_{tt}\equiv \partial E_t/\partial \tau_t$ ranges from $2$ to $7.5$, while
$C_{t1}$ ranges from $1$ to $2$ (except in the low tail of $\tau_1$). 

Summing up, bulk contacts play the leading role in determining the thermodynamical
properties of the designed sequence.
This is not only because there are more bulk (40) than interface (12) contacts. In fact,
the specific heat {\it per contact} is $C_{11}/40\approx 0.2$ for the bulk and
$C_{tt}/12\approx 0.8$ for the interface. Consequently, each interface contact is
more sensitive to the evolutive pressure (i.e. the evolutive temperature)
excerted on it than a bulk contact, not because they are less but because of their
different topology (see next Section).

Moreover, a relation similar to the Fluctuation--Dissipation theorem holds
\begin{eqnarray}
<\epsilon_1^2>-<\epsilon_1>^2&=&\tau_1^2 C_{11}/40; \nonumber\\
<\epsilon_t^2>-<\epsilon_t>^2&=&\tau_t^2 C_{tt}/12,
\end{eqnarray}
which links the energy fluctuations (l.h. side) to the specific heat (r.h. side). 
This is a consequence of the fact that the distribution probability (\ref{dis_dim}) is separable in
$E_1$ and $E_t$. Consequently, for values of $\tau_1$ and $\tau_t$ similar and in keeping that $C_{tt}/12>C_{11}/40$, one can conclude that each interface contact fluctuates more than a bulk contact.

\section{Consequences of the different topology of bulk and interface contacts}

The two main properties the space of sequences of dimeric proteins display are: a) 
it is not possible to optimize both the bulk and the interface contacts at the same 
time and b) the thermodynamical quantities characterizing the designed sequences are more
sensitive to the evolutionary pressure associated with $\tau_1$ (bulk) than to that associated with $\tau_t$ (interface).

Property a) is typical of all systems with disordered interactions. Like in
spin glasses, the disorder of the matrix elements makes it impossible to optimize
all parts of the system at the same time (cf. e.g. \cite{wolynes,parisi}).
This also happens in the case of monomeric, single--domain proteins, although one does not, as a rule, discriminate between the different contacts, imposing the same selective pressure
(i.e. evolutionary temperature) on all the residues.

The asymmetry between the behaviour of the bulk and of the interface is due
not only to the fact that there are more bulk than interface contacts, but also
because the topology of such contacts is very different.
In fact, the bulk is a three--dimensional system, composed of a rich network of interconnections (cf. Fig. 1(b)), so that a residue can be correlated with
other residues leading to  long range order. On the other hand, the interface is composed of contacts which are essentially
independent of each other, although they are not independent of the bulk contacts. (cf. Fig. 1(c)).
This implies that, in principle, it is easier for an interface contact to be at a low energy
than for a bulk contact, in keeping with the fact that it is easier to optimize an uncorrelated system than
a correlated one. But it also means that the interface is less stable then the
bulk, the associated native contact energies displaying much stronger variations than that associated with bulk contacts.

 The basic difference existing between bulk and interface contacts can be further clarified by comparing the thermodynamics of
a typical lattice model protein (in this case a 36-mer) and a system composed
of the same number of contacts, but placed, in the lattice, independently of each other \cite{derrida}.
In Fig. 6 is displayed the average energy for the protein contacts 
and for the independent contact system
(solid and dashed curves, respectively). The average energy, sum of uncorrelated contributions, can reach an
average value which is, at any design temperature (evolutionary pressure), lower than the energy of the bulk. 
The shape of the energy function is well described by the predictions of the
Random Energy Model \cite{derrida}, i.e. $<E>\sim -T^{-1}$ (dotted curve in Fig. 6).
On the other hand, the specific heat of the independent--contact--system (i.e. the slope of 
the dashed curve) is always larger than that of the bulk, 
indicating that the fluctuations in the
independent--bond--system which, through
the Fluctuation--Dissipation theorem, are proportional to the specific heat, are larger than in the bulk. In other words,
independent contacts can reach lower energies but are more unstable than correlated
contacts.

The different thermodynamics associated with the interface and with bulk contacts is also reflected by the different conservation 
patterns for the amino acids lying on the interface and in the bulk.
Because three--state dimers concentrate their energy in the highly interconnected bulk, they
display a conservation pattern typical of monomeric proteins in which
few sites are higly conserved while the remaining sites can mutate more
freely \cite{jchemphys1}. If one calulates for each site its sequence entropy 
$S(i)=-\sum_{\sigma=1}^{20} p_i(\sigma)\log p_i(\sigma)$, where $p_i(\sigma)$
is the probability to find the amino acid of kind $\sigma$ in the $i$th site, then 
the resulting distibution of $S(i)$ displays a low--S tail (cf. Fig. 7(b)).

On the contrary, in the case of two--state dimers, the energy is concentrated
on the interface. Since the interface contacts are independent on each other,
none of them is privileged, so that the degree of conservation is more uniformly distribuited among a large number of amino acids than before. 
The resulting distribution of $S(i)$ is that shown in Fig. 7(a)
and displays a sharper behaviour than that associated with three--state dimers. 
These results are in overall agreement with the findings 
by Grishin and Phillips, who analyze the conservation of residues on the 
surface of five two--state dimeric enzymes and find no signal of any larger conservation
on the interface than in the bulk \cite{grishin}.

The different topologies also imply that while the
stabilization energy of three--state dimers is concentrated in few sites buried in
the bulk and a mutation of one of these "hot" sites causes misfolding of the
protein \cite{jchemphys1}, in two--state dimers the energy is distributed
more evenly on the interface, so that its sites are more tollerant to mutations.
This allows two--state dimers to build active sites on the interface, for which
purpose the protein has to mutate stabilizing residues with residues which
perform other biological tasks.

\section{Folding and aggregation}

The folding mechanism of three--state homodimers is, within the lattice model, 
essentially the same as that of monomeric proteins, with the additional association
step. First, LES are stabilized by strongly interacting residues
which are close along the chain (the conserved residues discussed above). When the LES assemble together to form the folding nucleus, the protein
folds to its monomeric native state \cite{jchemphys2,jchemphys3}. The time limiting step
is the association of the two monomers into the dimers, which is controlled both by
the diffusion constant and by the stability of the interface.

The behaviour of two--state dimers is different. First the interface is built, which
is the time limiting step, and then the rest of the protein folds around the
interface. This is again a nucleation event, which causes a sudden transition from
the unfolded state to the dimeric native conformation, but involves a larger number
of residues than in the case of three--state dimers. This mechanism is compatible with
the finding that two--state dimers, like arc repressor \cite{sh_nsb}, display small
$\varphi$--values (which is defined as the relative change in free energy between the native conformation and the transition state ensemble upon mutation \cite{fersht}). In fact, since the transition state is determined by the association
of the dimers and the residues at the interface share evenly their stabilization
energy, the free energy difference between native and transition state is also
distributed among a large number of residues. In other words, none of the interface
residues has a leading role in the formation of the interface.

Sequences for which the strong evolutionary pressure on the interface is
not balanced by a strong pressure on the bulk display specific aggregation. The term specific means that, although unstructured, the aggregate display some recursive interactions, typically between residues belonging to the interface. For example, in the case of sequence S$_5$ listed above, that between residue 3 of one monomer and 6 of the other monomer.
In the design of this sequence the low value of $\tau_t$ (strong evolutionary pressure) 
selects for the interface a subset of the 20 kinds of amino acids with quite low
average contact energy (their average value being ${\overline B}'=-0.12$, as compared to a zero value for the MJ contact energies, with standard deviation $\sigma_B'=0.32$, to be compared with $\sigma=0.3$ for the MJ contact energies \footnote{This result also implies a quite low local value of the threshold energy $E_c=NB'-N\sigma_B'(2\ln\gamma)^{1/2}$ (cf. text). In fact, assuming it to be valid for both the interface and the bulk, one would obtain (making use of $\gamma=2.2$, which for ${\overline B}=0$ and $\sigma_=0.3$ gives $E_c=-35$) $E'_c=-48.3$.}). A concentration of such strongly interacting residues makes it easy the assembly of the interface in a number of ways which are different from that of the native conformation, in a similar way as if a monomeric protein were composed only of strongly interacting residues, it would display a miriad of low energy conformations competing with the native one. To be more precise,
the native interface ($E_t=-5.11$) has to compete with other conformations
that the first twelve residues of each chain can assume, conformations which have
energies of the order of $12{\overline B}'-12\sigma_B'(2\log\gamma)^{1/2}=-6.2$
(again evaluated in the approximation of the random energy model \cite{derrida}), thus energetically more favourable than the native interface. Conversely,
the bulk is not so well optimized ($E_1\sim -15$) as to re--establish the energetic balance in favour of the native state. The outcome is a globule containing the first, strongly interacting,  twelve residues of each chain, surrounded by a disordered cloud made of the other monomers. 

This kind of aggregation is different from the aggregation of poorly designed sequences. In this case, the energy of the native state is quite high and more homogeneously distributed among the contacts, and the resulting equilibrium conformation is an ill defined clump stabilized by random interactions.

All the results described above have been found keeping fixed the size ($L=7$) of the cell which contains the system. The dependence of the behaviour of the system on this size, which reflects the concentration of monomers, is described in the Appendix.

\section{conclusions}

It has been shown that the bulk and the interface contacts contribute differently to
the folding mechanism of dimers, due to their different topology. 
That is, the bulk displays a rich network of contacts, while pairs of residues belonging to different monomers interact independently of each other.
This difference manifests itself
in the different conservation patterns expected in the case of two--state and three--state dimers, conservation patterns which are the main point of contact between model predictions and real proteins.

\appendix
\section{}

The folding behaviour dependens not only on the energetics of the sequence, but also on the monomer concentration $\rho$,
a quantity which is reflected, within the present model, by  
the linear dimension $L$ of the Wigner cell where the calculations 
are carried out through the expression $\rho=2/L^3$.

For $L<6$ the chains experience excluded volume violations. We found that
even at $L=6$ the chains get entangled at any temperature, never finding
the native conformation. This is due both to dynamical reasons, because of the lack
of space available to the movements, and thermodynamical ones, since each lateral sites
of the native conformation interact with those of a (virtual) neighbouring dimer,
interaction which was not optimized in the design process and, consequently, rises
the total energy of the system above $E_c$. 

For large $L$ the dimeric native state becomes unfavourable with respect to states in
which the chains are separated, due to the contribution of traslational entropy.
In the case of sequence $S_{2-state}$, the dimeric native state "$N_2$" has to compete with the state
in which the two chains are disjoint and folded in their monomeric conformation ("$2N_1$").
The free energy $F_N$ of the dimeric native state is equal to its internal energy
$E_N=-36.8$ (the state is unique, so its entropy is zero). The free energy of the state
$2N_1$ is $2E_1-TS_{trans}$. The translational entropy $S_{trans}$ is the logarithm
of the number of different conformations in which the Wigner cell of size $L$ can
accomodate the two native conformations in such a way that they do not interact,
that is $S_{trans}=\ln [(6\cdot 4 ((L+1)^3-(c+1)^3)]$, where the numeric factors
takes into account the different orientation of the monomers once their centre of
mass is fixed, the square parenthesis indicate the number of possible positions of
the centre of mass, in the approximation that the native monomer is a cube of length $c$
(in the present case $c$ is, in average 3.3).
The critical size $L_c$ of the Wigner cell, which separates the regimes of
dominance of the state "$N_2$" rather than "$2N_1$" (which corresponds to the critical
concentration $\rho_c=2/L_c^3$) is obtained equating the associated free energies, which
gives
\begin{equation}
\label{lc}
L_c=\left[(c+1)^3+\frac{1}{24}\exp\left(-\frac{E_t}{T}\right)\right]^{1/3}-1.
\end{equation}
In the case of sequences of three--state dimers like $S_{3-state}$, which have $E_t$ in the
range $-3/-4$ one obtains from Eq. (\ref{lc}) $L_c$ ranging from 12 to 30 at $T=0.28$, 
at which the folding
is fastest and which regard as "room" temperature (cf. \cite{jchemphys3}). At values of $L$
larger than $L_c$ each monomer, which in the native state has an energy $E_1\ll E_c(N=36)=-14$
is stable (typically $E_c(N=36)-E_1$ is of the order of 10$T$).

For two--state dimers the critical size of $L_c$, calculated with Eq. (\ref{lc}), 
is of the order of $150$, much larger than the critical size of three--state dimers. Moreover, even if one choses a low concentration (large $L$) such that
the dimeric native state becomes unstable, each monomeric native state ($N_1$ state)
has an energy larger than the case of sequences of kind $S_{2-state}$, i.e. of the order of
few--$T$, so that it is quite unstable, having to compete with the sea of monomeric 
unfolded conformations. 

In the calculations of the phase diagram of Fig. 2, we have chosen $L=7$, a choice which assures
stability of all sequences selected and, as discussed above, allows the movement
of the chains. 
Even if with this choice, the system experiences some difficulties in 
reaching the native conformation,
due to the narrowness of the space available. For example sequence $S_{2-state}$ can find its correct dimeric native state 
in 16 times out of 20.
In fact, in 4 cases it finds a conformation with energy $E=-35.92$,
corresponding to a situation in which one of the two chains (let us call them A and B)
is folded (say, chain A),
the monomers of chain B at the interface being in their native position, while chain 
B is in a (well
defined) conformation which has only $40\%$ similarity with its native structure
(similarity parameter $q=0.4$). The reason for this result is to be found in the fact that
chain B builds some contacts with the "back" of chain A, taking advantage of the periodic
boundary conditions. These contacts are mostly between residues of chain B and partners which are
of the right kind but belong to the wrong chain (contacts 17A--32B, 23A--18B, 24A--17B, 25A--36B,
26A--35B, 35A--26B), while two of them are between monomers which cannot be in contact if they
belong to the same chain (31A--32B, 32A--33B). 

Such conformation is a metastable state which, even slowing down the folding process,
does not interfere seriously with the thermodynamics of this two--chain model. The
situation is different if one consider a more realistic system built of a number of proteins
comparable to the Avogrado's number.
In this case, the system can build a chain of dimers which gain an additional energy
hrough this back--binding. This could give rise to regular structures resembling amyloid
fibrils.



\newpage

\begin{figure}
\label{fig1}
\caption{(a) The native conformation used in the present calculations. (b) The rich network of 40 contacts formed by the residues belonging to one of the monomers displayed in (a). (c) The 12 contacts between the two monomers.}
\end{figure}

\begin{figure}
\label{dyn_phases2}
\caption{Dynamical phase diagram of dimeric sequences selected at different values
of the evolutionary temperatures $\tau_1$ and $\tau_t$. Solid squares indicate 
two--state folding sequences, solid triangles three--state folding sequences,
empty symbols label sequences which display specific (diamonds) and unspecific (circles)
aggregation. The solid curve corresponds to the loci of the set of values of $(\tau_1,\tau_t)$ associated with the
total energy $E_c$, while dashed curves limits the area outside which  the specific heat associated with either the volume or the interface degrees of freedom is
negative.}
\end{figure}

\begin{figure}
\caption{Average energy per contact associated with the bulk ($2E_1/80$), with the interface ($E_t/12$) and with the energy of the two identical designed monomer in the native conformation ($E_{desig}/92=(2E_1+E_t)/92$ of the dimer whose native conformation is shown in Fig. 1(a). Also shown are the results associated with the isolated monomer S$_{36}$. Also reported is the average energy per contact associated with the threshold energy $E_c$ associated with both the dimer ($=-35$) and the monomer (S$_{36}$, $E_c=-14.5$). In parenthesis we display  the differential variation (in pergentage) of the quantities associated with two consecutive sequences from left to right. }
\end{figure}

\begin{figure}
\label{entropy}
\caption{The entropy of the space of sequences as function of $E_1$ and $E_t$. The
lowest energy states with respect to $E_1$ and $E_t$ are indicated. The isoentropic curves which separates the grey levels correspond to variations in the entropy of 20 (in the same dimensionless units used also for the energy). The darker areas correspond to lower values of the entropy.}
\end{figure}

\begin{figure}
\label{fig4}
\caption{The average energy $E_1$ (a) and $E_t$ (b) as function of $\tau_1$ and $\tau_t$.}
\end{figure}

\begin{figure}
\label{fig5}
\caption{The average energy as a function of evolutionary temperature for a monomeric model protein
of length 36
(cf. ref. \protect\cite{jchemphys3}, solid line) and for a system of 36 independent bonds (dashed line). The monomeric protein displays the random energy behaviour $<E>\sim \tau^{-1}$ at high evolutionary temperature and a linear behaviour al low temperature, while the independent bond system displays an almost perfect random energy behaviour (dotted line). }
\end{figure}

\begin{figure}
\label{fig6}
\caption{The distribution $P(S)$ of entropy per site,calculated for (a) sequence $\#1$ (two--state dimer), (b) sequence $\#3$ (three--state dimer),
(c) sequence $\#5$ (aggregation) and (d) sequence $\#6$ (monomer S$_{36}$).}
\end{figure}

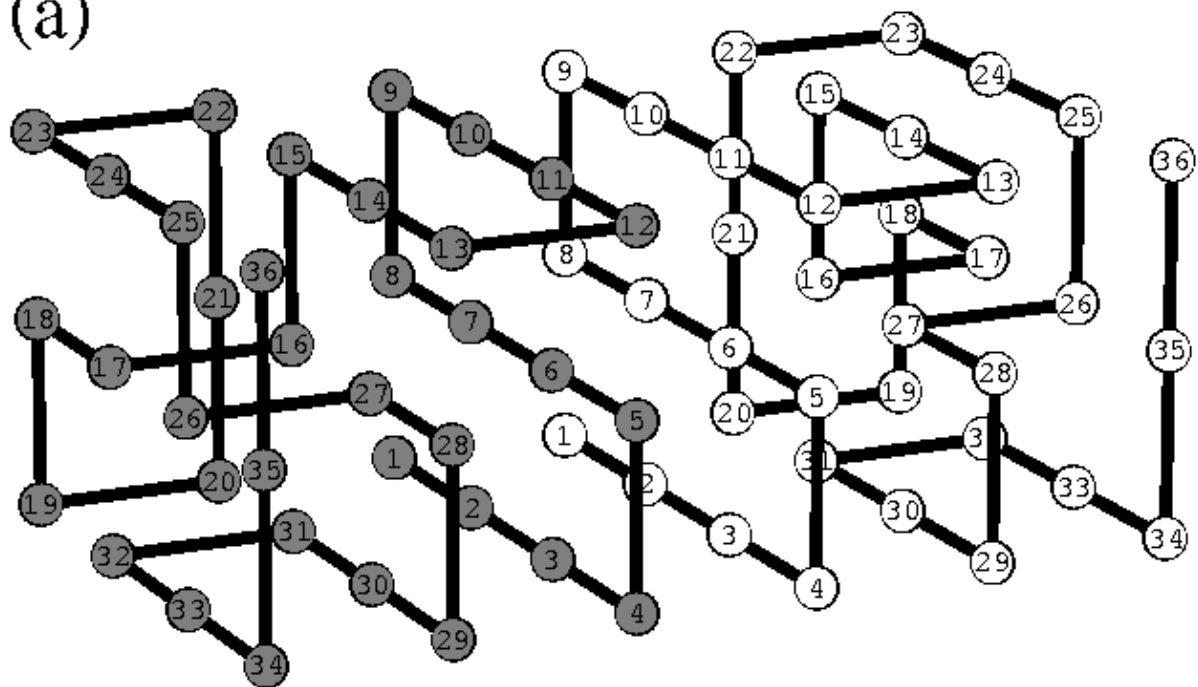
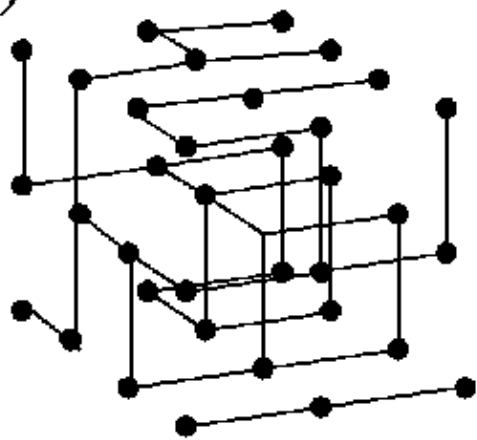
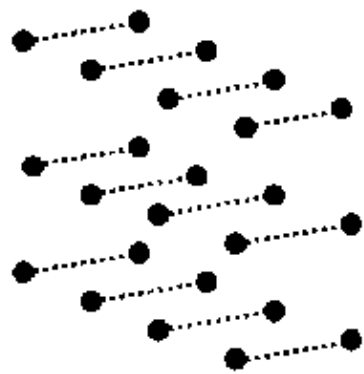
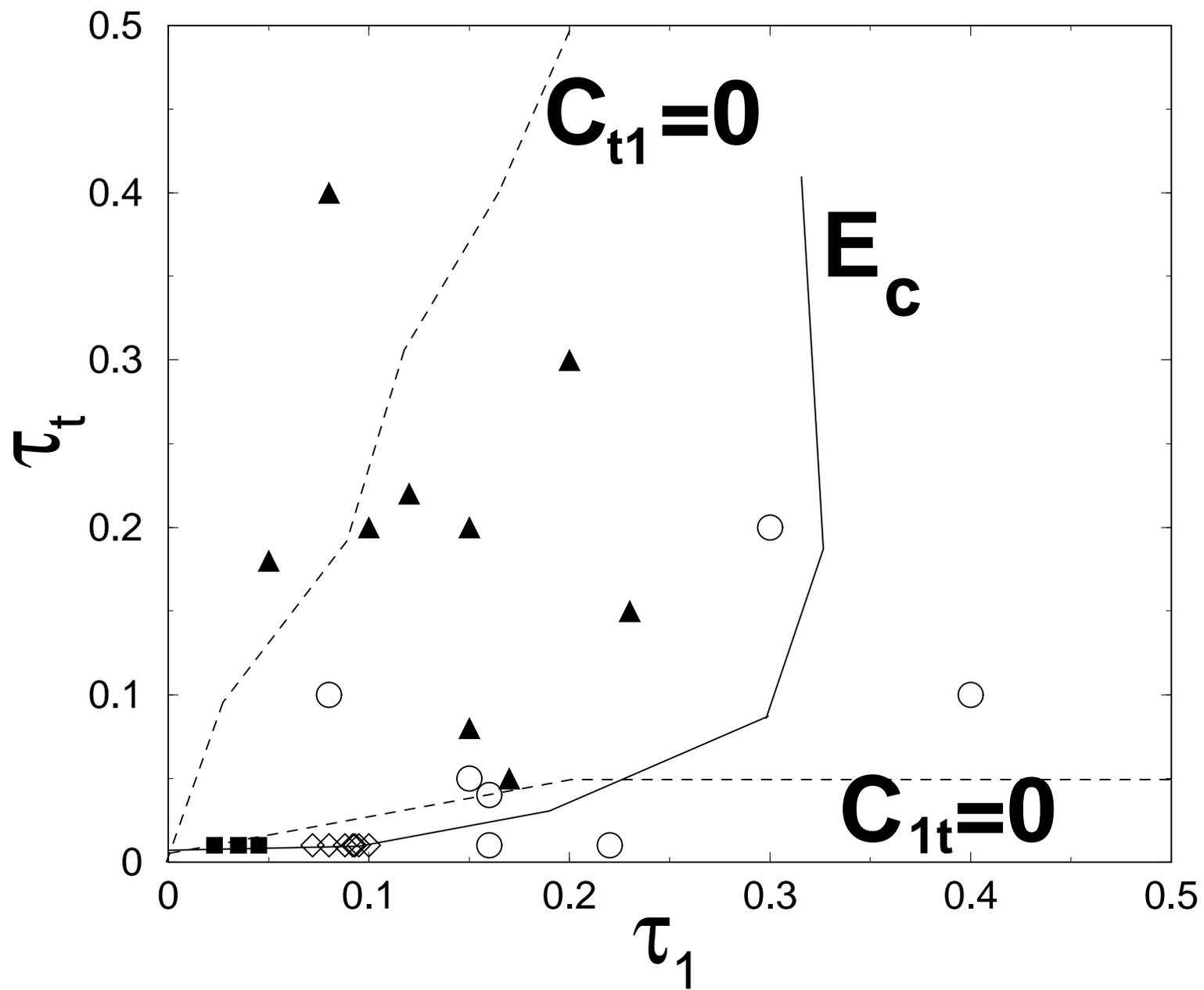
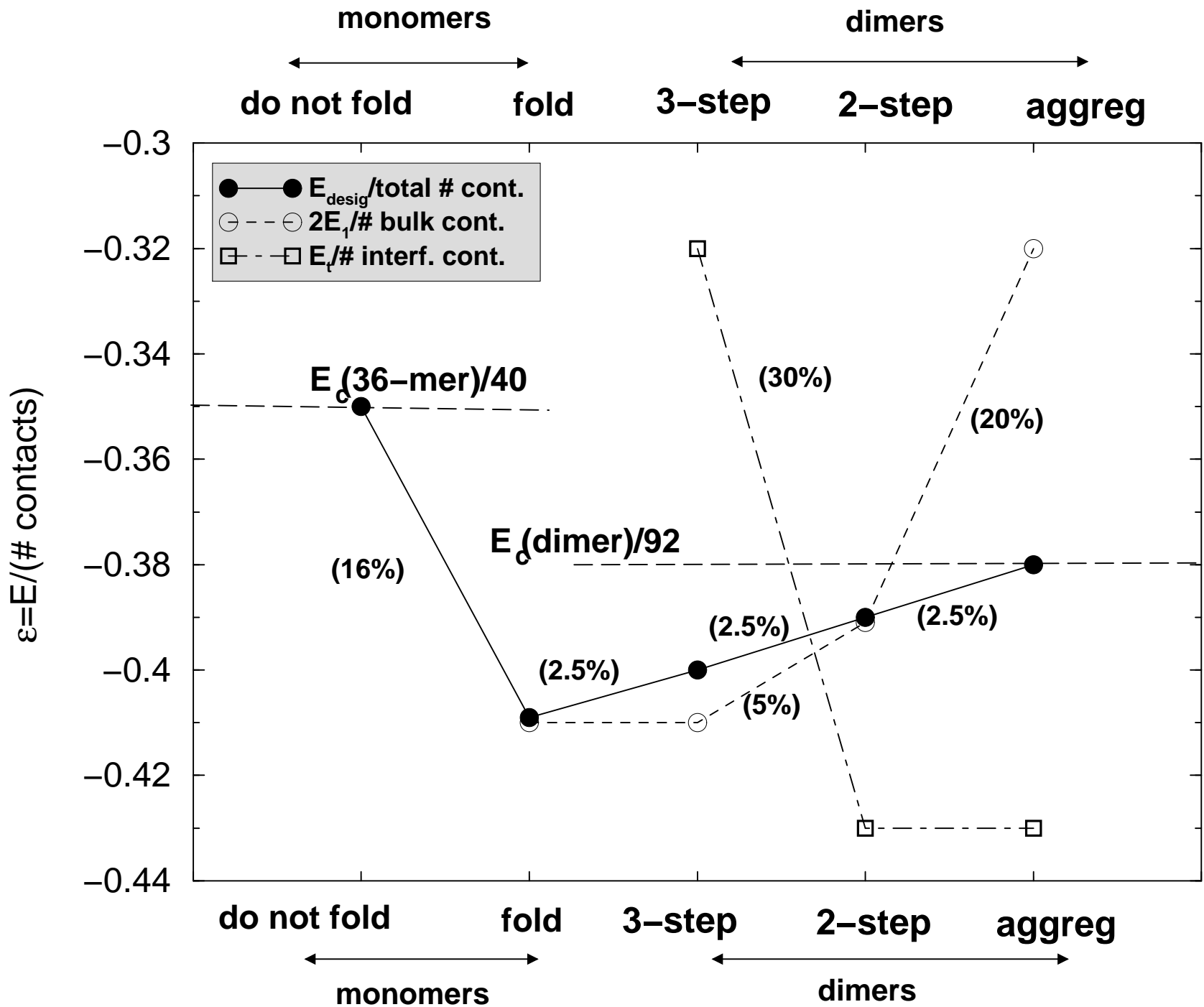
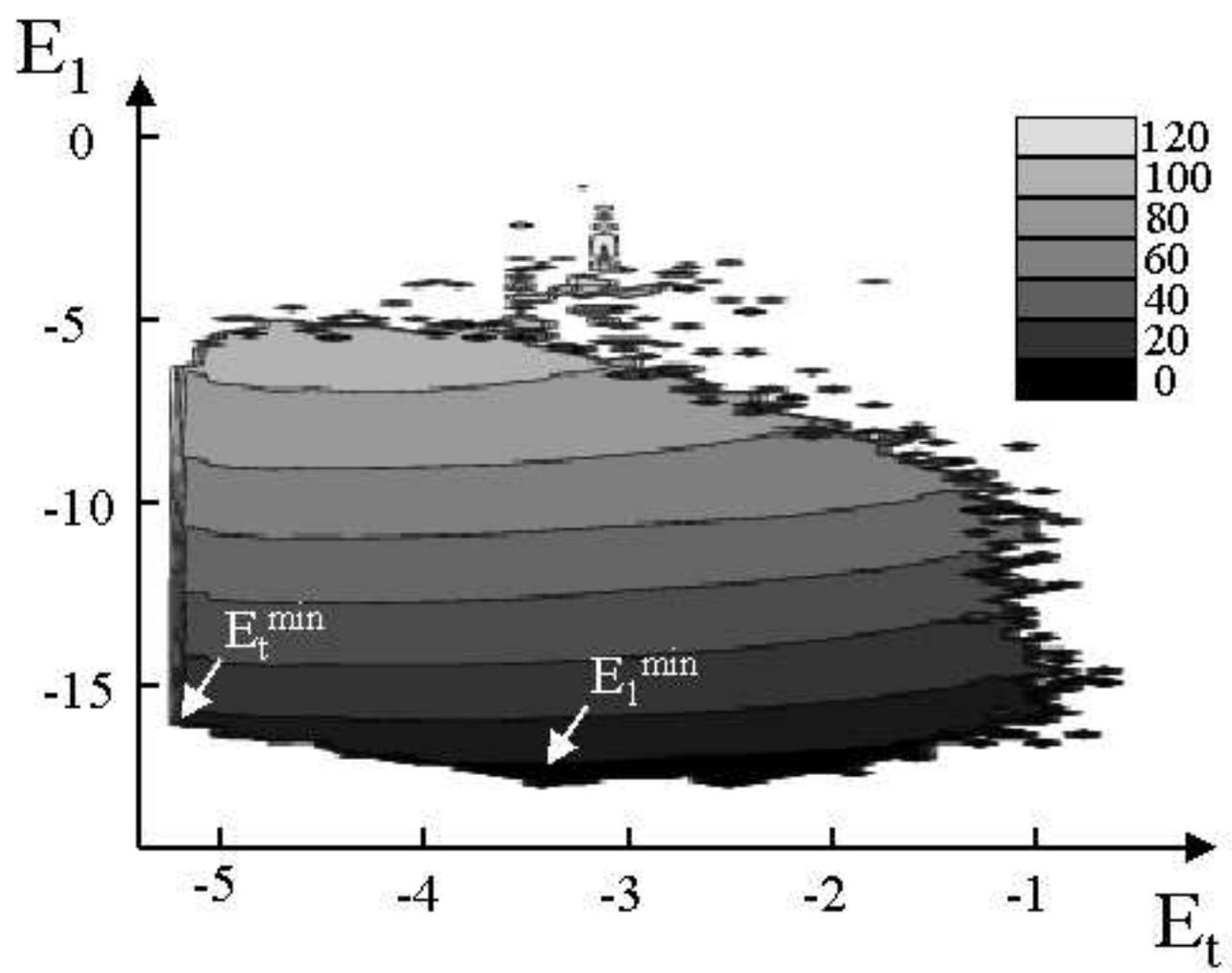
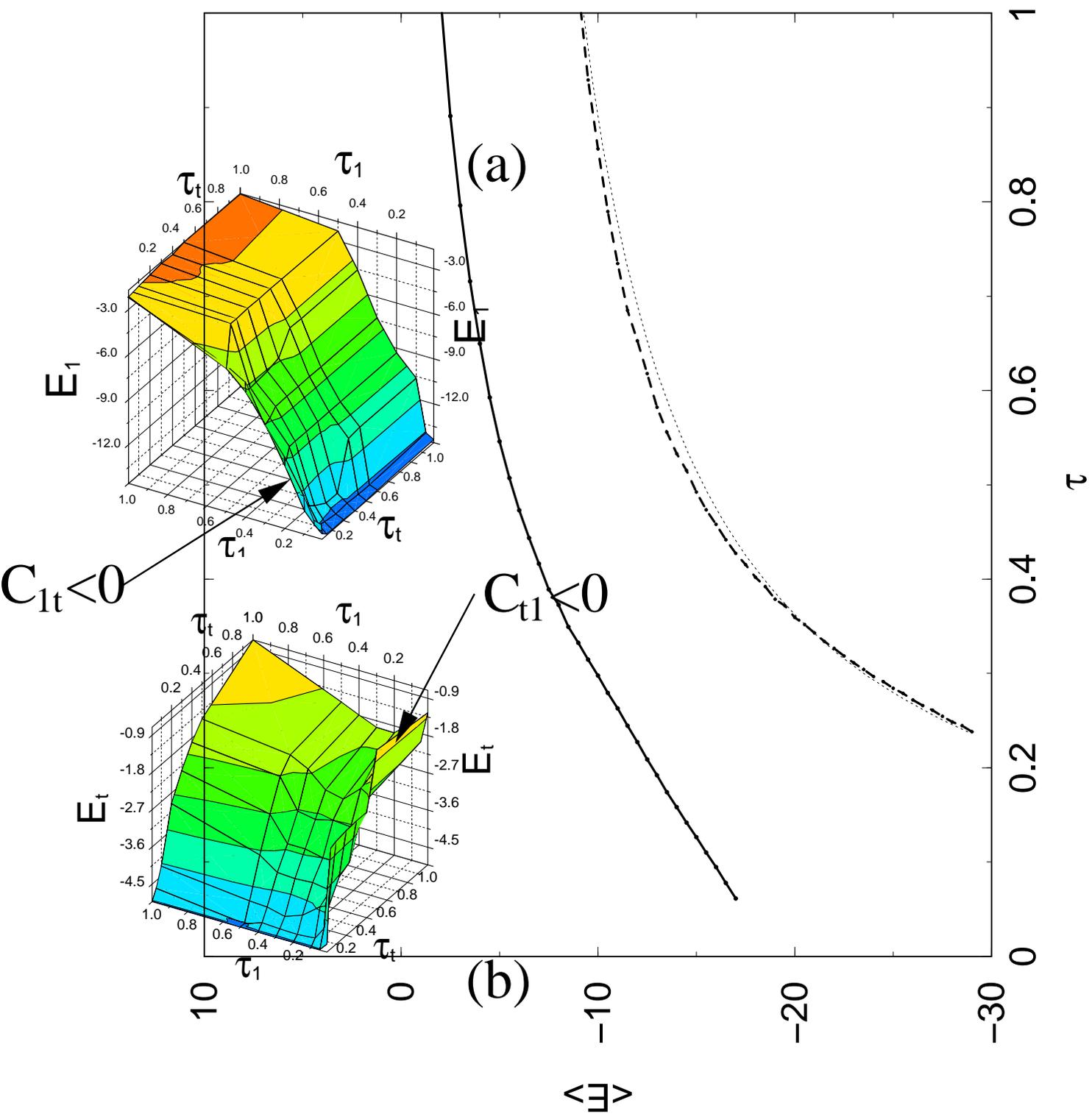
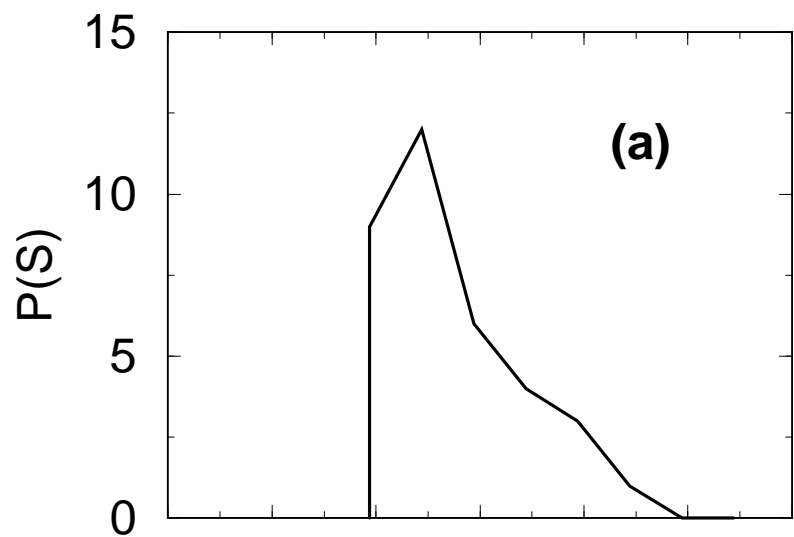
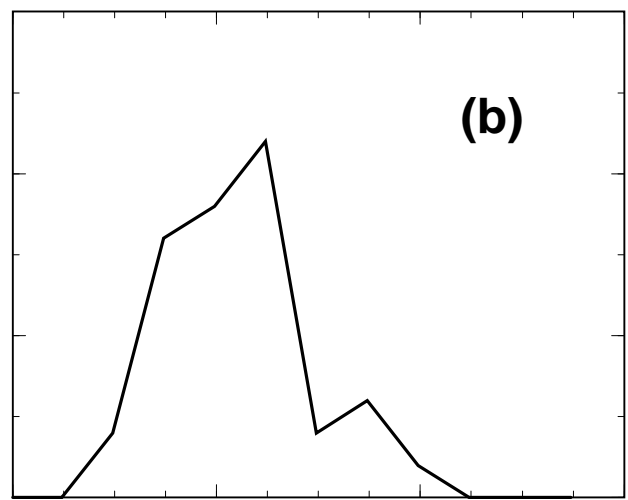
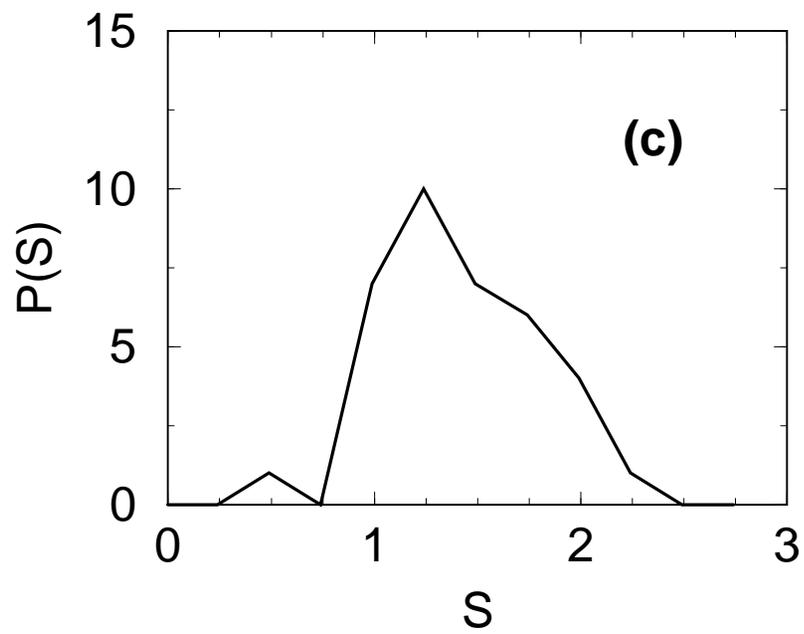
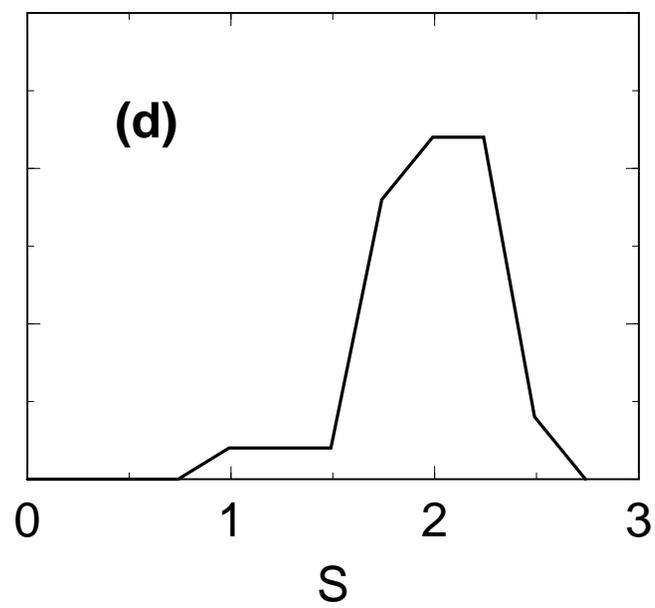
\end{document}